\documentclass[10pt,conference]{IEEEtran}
\usepackage{cite}
\usepackage{amsmath,amssymb,amsfonts}
\usepackage{algorithmic}
\usepackage{graphicx}
\usepackage{textcomp}
\usepackage{xcolor}
\usepackage{tcolorbox}
\usepackage{booktabs}
\usepackage[table]{xcolor}
\usepackage{makecell}
\usepackage{geometry}
\usepackage{lipsum}
\usepackage{accessibility}
\usepackage{enumerate}

\begin{document}

\newcommand{\red}[1]{\textcolor{red}{#1}}

\newcommand{\RQone}{
How do the types of code review comments written by humans differ from those generated by an LLM?
}

\newcommand{\RQtwo}{
What types of LLM-generated code review comments are frequently resolved by developers?
}

\title{What Types of Code Review Comments Do Developers Most Frequently Resolve?}

\author{
\IEEEauthorblockN{
Saul Goldman\IEEEauthorrefmark{1}, 
Hong Yi Lin\IEEEauthorrefmark{1},
Jirat Pasuksmit\IEEEauthorrefmark{2},
Patanamon Thongtanunam\IEEEauthorrefmark{1}, 
Kla Tantithamthavorn\IEEEauthorrefmark{2}\IEEEauthorrefmark{4},
Zhe Wang\IEEEauthorrefmark{2}, \\
Ray Zhang\IEEEauthorrefmark{2}, 
Ali Behnaz\IEEEauthorrefmark{2}, 
Fan Jiang\IEEEauthorrefmark{2}, 
Michael Siers\IEEEauthorrefmark{2}, 
Ryan Jiang\IEEEauthorrefmark{2}, 
Mike Buller\IEEEauthorrefmark{2}, 
Minwoo Jeong\IEEEauthorrefmark{3}, 
Ming Wu\IEEEauthorrefmark{3}
}
\IEEEauthorblockA{\textit{
    \IEEEauthorrefmark{1}The University of Melbourne, Australia, \IEEEauthorrefmark{2}Atlassian, Australia. 
    \IEEEauthorrefmark{3}Atlassian, USA. 
    \IEEEauthorrefmark{4}Monash University, Australia. 
}}
}

\maketitle

\begin{abstract}
Large language model (LLM)-powered code review automation tools have been introduced to generate code review comments.
However, not all generated comments will drive code changes.
Understanding what types of generated review comments are likely to trigger code changes is crucial for identifying those that are actionable.
In this paper, we set out to investigate (1) the types of review comments written by humans and LLMs, and (2) the types of generated comments that are most frequently resolved by developers.
To do so, we developed an LLM-as-a-Judge to automatically classify review comments based on our own taxonomy of five categories.
Our empirical study confirms that (1) the LLM reviewer and human reviewers exhibit distinct strengths and weaknesses depending on the project context, and (2) readability, bugs, and maintainability-related comments had higher resolution rates than those focused on code design.
These results suggest that a substantial proportion of LLM-generated comments are actionable and can be resolved by developers.
Our work highlights the complementarity between LLM and human reviewers and offers suggestions to improve the practical effectiveness of LLM-powered code review tools.
\end{abstract}



\section{Introduction}
Code review is a standard practice in modern software development, with developers typically spending 10–15\% of their time on this task~\cite{BosuCBOC17, thongtanunam2014improving,thongtanunam2022autotransform}.
It plays a vital role in enhancing code quality and promoting team collaboration~\cite{RigbyB13,McIntoshKAH16}.
A typical code review workflow requires a developer other than the code author to review and provide feedback for new code changes before merging into the main codebase.
Despite its benefits, code review can be time-consuming and demands considerable effort from human reviewers~\cite{BaumLNS17,GoncalvesFBSB22}. 
To alleviate this burden, Large Language Model (LLM)-powered code review tools have been developed
to automate various code review activities (e.g., generating code review comments, code refinement)~\cite{codereviewer}.

Review comment generation plays a vital role in code review automation as comments can trigger various improvements, contributing varying levels of value to the code review process.
To better understand which types of code review comments are being generated, researchers  manually classified a generated comment into a category~\cite{tufano22, lin24}.
Understanding this is crucial for advancing LLM-powered code review tools and guiding product design and development, which subsequently enhances code review productivity.
In particular, when comments are routinely accepted or resolved with code changes, it demonstrates strong alignment with developers’ needs and workflows, thereby driving tool adoption and amplifying their impact on developer productivity.

In this paper, we set out to investigate (1) the comment types generated by LLM reviewers compared to those provided by humans, and
(2) the types of LLM-generated comments that are most frequently resolved by developers.
To do so, we first developed a taxonomy of comment types based on five categories (i.e., readability, bugs, maintainability, design, and no issue).
Then, we developed an LLM-as-a-Judge to automatically categorize review comments based on the aforementioned taxonomy.
After confirming the Judge's reliability, we categorized both LLM-generated and human-written comments, enabling us to analyze the differences in issue types discussed.
Finally, we classified comments that were previously generated for internal projects at Atlassian to facilitate a triangulated evaluation of resolution rate by comment type. 
Then, we answer the following RQs:

\begin{enumerate}[{\bf RQ1)}]
    \item {\bf \RQone} 
    For Atlassian projects, the LLM reviewer generated a higher proportion of comments related to bugs and maintainability, but fewer comments focused on readability compared to human reviewers.
    For OSS projects, the LLM reviewer generated more maintainability-related comments but fewer design-related comments than human reviewers did.
    Nonetheless, bug-related comments are less-frequently generated by an LLM reviewer.
  \item {\bf \RQtwo}
  Readability, bug, and maintainability-related comments exhibited higher resolution rates compared to those focused on code design. 
  Nevertheless, many of the LLM-generated comments are not resolved by developers. 
  Such unresolved comments could be due to the lack of clarity, relevancy, and simplicity of the comments.
  
\end{enumerate}

These findings lead us to conclude that the LLM reviewer and human reviewers exhibit distinct strengths and weaknesses depending on the project context.
While LLM-generated comments about readability, bugs, and maintainability are more likely to be addressed by developers, bug-related comments remain underrepresented in LLM comments.
Balancing the distribution of comment types and improving the quality and clarity of LLM-generated comments is therefore crucial for increasing their  impact in code review workflows.

\section{Background and Related Work}

\textbf{Review Comment Categorization.}
Various taxonomies of comment types have been developed based on human-written review comments~\cite{MantylaL09,BellerBZJ14, TurzoB24}.
In general, code reviews discuss either functional or evolvability issues.
The former encompasses defects that can cause system failures at execution time (e.g., incorrect implementation, logical errors, mistiming), while the latter involves issues that affect the compliance, maintainability and understandability of the codebase (e.g., documentation, refactoring, code readability)~\cite{MantylaL09}.
To understand benefits of code reviewing, prior studies explored how different types of human-written comments relate to software quality~\cite{ThongtanunamMSR2015} and developer-perceived usefulness~\cite{BosuGB15, HasanIIRB21}.
Despite extensive research on the comment types in human-written comments, the differences in their distribution compared to those in generated comments remain largely unexplored.

\textbf{Code Review Comment Generation and its Evaluation}
Recent studies have explored automating review comment generation to reduce the manual effort involved in code reviews. Many approaches use sequence-to-sequence models—such as the original Transformer~\cite{tufan2021towards}, T5~\cite{tufano2022using}, and CodeT5~\cite{tufano2024code}—or leverage large language models (LLMs) like LLaMA~\cite{llamareviewer}, GPT-3.5~\cite{pornprasit2024fine}, and GPT-4o~\cite{olewicki2024impact}. 
Generated comments are typically evaluated against human-written references using surface-level similarity metrics such as BLEU, ROUGE, and Exact Match, which measure lexical similarity.
However, these metrics overlook the diversity and semantics. 
To address this, recent work~\cite{tufano22, lin24, tufano21, tufano_strengths,hong2022commentfinder} have conducted manual classification of comment types to further evaluate the generated review comments.
Although recent work has explored the use of an LLM as a judge to classify code review comments, the primary focus has been on curating datasets for training purposes~\cite{curev}.
Despite the valuable insights of comment types, little is known about which types of generated comments developers actually resolve in practice.

\section{Research Methodology}


In this section, we present the motivation of our research questions and the research methodology.


\subsection{Research Questions}

\noindent \textbf{RQ1: \RQone}  

\noindent\underline{Motivation.} 
Investigating differences in the distribution of comment types between human-written and LLM-generated code review comments can reveal the distinct strengths and limitations of LLMs in reviewing code changes. 
By identifying these differences, we gain insight into the comment types that LLMs tend to prioritize and  their behavior aligns with the code review practices of human reviewers.

\noindent \textbf{RQ2: \RQtwo}  

\noindent\underline{Motivation.} 
While LLMs can generate a wide range of review comments, it remains unclear which comment types are more likely to lead to comment resolution (i.e., the corresponding code line is changed in the subsequent commit).  
Understanding the comment types that are most often resolved by developers is crucial for understanding the impact of LLM-generated comments in real-world use cases.
Such insights will help inform future efforts to enhance the effectiveness of LLM-powered code review tools.

\subsection{A Taxonomy of Comment Types}
\begin{table*}[t]
\caption{Our Taxonomy of The Types of Code Review Comments.}
\centering
\label{tab:categories}
\rowcolors{2}{white}{gray!10} 
\renewcommand{\arraystretch}{1.5}
\begin{tabular}{p{2cm} p{13.6cm}}
\Xhline{2\arrayrulewidth}
\rowcolor{gray!40} 
\textbf{Categories} & \textbf{Description} \\ \hline 
\rowcolor{gray!10} 
Code Readability & Focuses on the textual readability of code that does not impact functionality.
Includes naming conventions, syntax, formatting, code duplication, unused variables, etc. 
Excludes issues related to code comments or documentation. \\

Code Bugs & Identified faults in software functionality that caused implementation incorrectness. 
Includes misuse of language features, logical errors, incorrect variable types, validation issues, resource management, timing problems, and interface interactions. \\

Maintainability & Addresses issues or suggests changes related to the maintainability of the system and the need for proper code comments and documentation.
Includes concerns on hard-coding, fragile code, configuration, best-practices, logging, future-proofing, modifying or adding tests, performance optimization, cyclic dependencies, security, DevOps, version control (e.g., git), etc. \\

Code Design & Concerns regarding code complexity and structure, appropriate implementation patterns, design principles, and architecture.
Includes refactoring, code organization, state management, model structure, null handling, additional programmatic checks, etc. \\

No Issue & Not directly relevant to code readability, code bugs, maintainability or code design. 
Focuses on raising open-ended discussions, affirmation, humor, or acknowledgment. \\

Other & Does not fit into the other categories. \\ \Xhline{2\arrayrulewidth}
\end{tabular}
\end{table*}
To better understand the types of code review comments, we constructed a taxonomy of comment types that are mostly aligned with human-written code review comments at Atlassian.
We began with a random sample of 336 human-written comments from the merged pull requests.
To ensure the quality of review comments, we selected only comments longer than 20 characters, inserted to a line in a source code or test file (excluding auxiliary files such as compiled bytecode, dependency locks, or static assets), and not written by the pull request author or automated bots~\cite{lin24}.

To construct a taxonomy, six Atlassian software engineers conducted card sorting on the sampled review comments to explore the types of code review comments.
The annotators iteratively refined the category definitions together until a consensus was reached.\footnote{As the card sorting and taxonomy development were done collaboratively by six engineers in a session, an inter-rater agreement could not be calculated.}
Table~\ref{tab:categories} presents the six comment types and their descriptions.

\begin{figure}[t]
    \centering
    \includegraphics[width=\columnwidth]{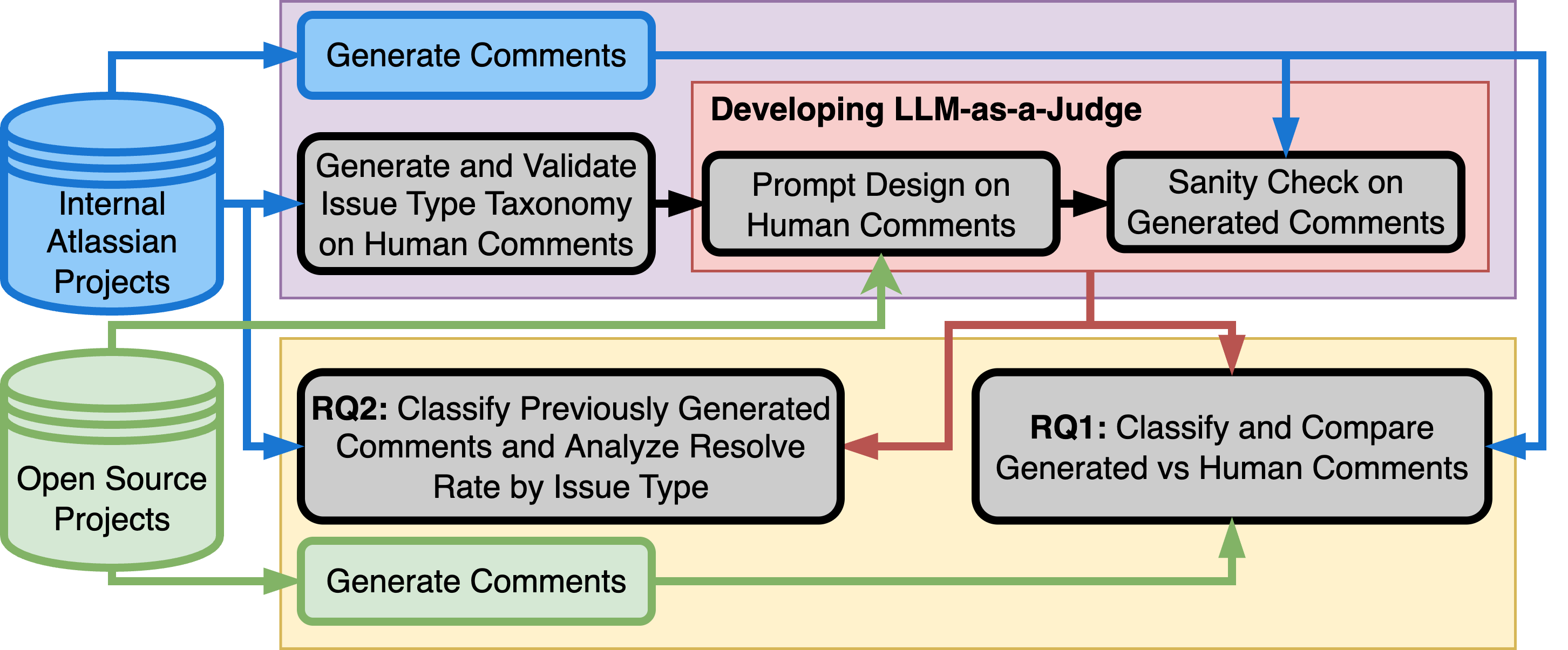}
    \caption{An overview of our research methodology to answer the research questions.}
    \label{fig:PD}
\end{figure}




\subsection{An LLM Judge for Review Comment Classification}
\label{sec:LLM-as-a-judge}
To classify code review comments using our taxonomy, we employed OpenAI's \texttt{gpt-4.1-2025-04-14} as an annotator.
As recent studies have demonstrated the effectiveness of LLMs-as-a-Judge~\cite{curev, takerngsaksiri2024human}, this approach will enable us to efficiently categorize a large volume of comments.
To ensure clarity and consistency, we designed prompts that are succinct, informative, and structured using a JSON format (Figure~\ref{fig:prompt-template}). 
Each prompt includes four components: an instructional message, a code review comment, the taxonomy, and an example response. 
We developed and refined the prompt iteratively based on human-written comments in open source projects~\cite{typescript}. 
The initial prompt simply instructed the LLM to select the most appropriate category based on the definitions without providing additional context.
To enhance performance, we introduced a justification prompt, encouraging the model to explain its classification. This implicitly triggers Chain-of-Thought (CoT) reasoning, which has been shown to improve LLM performance in classification tasks~\cite{deng_implicit_2023}. Including justifications also allows us to evaluate whether the model is reasoning about relevant features or guessing, thereby informing prompt refinement. We further extended the prompt to include a confidence score (ranging from 0 to 1) for each prediction, providing insight into how the model assesses its own ability to classify each individual code review comment.

\textbf{Sanity Check.}
To check the soundness of our LLM judge, we evaluated the method on 100 comments generated for our internal projects.
The comments were split into two subsets and independently labeled by two annotators over two rounds---near perfect agreement was achieved for each round (Cohen's $\kappa$ =\{0.80, 0.86\}).
After each set, they resolved all disagreements with a third annotator acting as an arbiter.
Finally, we use the full set of comments with human annotations that were agreed upon to measure inter-rater agreement with the LLM judge---moderate agreement was achieved (Cohen's $\kappa$ =\{0.42\}).
Corroborating with past studies~\cite{curev}, we deem the performance of the approach sufficient for evaluating code review comments.

\begin{figure}[htbp]
    \centering
    \input{Figures/prompt_template}
    \caption{The Main Prompt Template for Review Comment Classification.}
    \label{fig:prompt-template}
\end{figure}

\section{Results}

\subsection*{\textbf{(RQ1)\RQone}}
\begin{figure}[t]
    \centering
    \includegraphics[width=\linewidth]{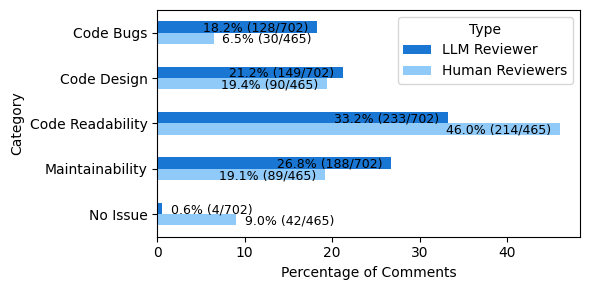}
    \caption{(RQ1-1) The distribution of human-written vs LLM-Generated code review comment types for Atlassian's internal projects.}
    \label{fig:rq2-part1}
\end{figure}

\begin{figure}[t]
    \centering
    \includegraphics[width=\linewidth]{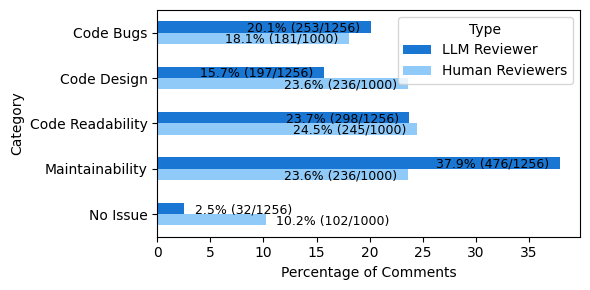}
    \caption{(RQ1-2) The distribution of human-written vs LLM-Generated code review comment types for OSS projects.}
    \label{fig:rq2-part2}
\end{figure}

\textbf{\underline{Approach.}} To answer this RQ, we aim to examine  differences in the distributions of comment types between LLM-generated and human-written comments.
To ensure that the results are not bound to specific contexts, we focus on two software development contexts, i.e., enterprise-graded internal software repositories at Atlassian and open-source software (OSS) projects.
For internal projects, we used 300 merged pull requests, containing 465 human-written comments.
For OSS projects, we used 1k human-written comments from the CuRev dataset---a curated code review dataset~\cite{curev}.
For each code change, we used RovoDev Agent (via Claude 3.5 Sonnet) to generate code review comments (i.e., an \emph{LLM reviewer}).
In total, the LLM reviewer generated 1,256 and 702 comments for open source and Atlassian internal projects, respectively. 
Note that the total number of generated comments are higher than the total number of human-written comments because the LLM reviewer autonomously generates comments for the pull-request.
Then, we used our LLM-as-a-Judge (see Section \ref{sec:LLM-as-a-judge}) to classify both human-written and LLM-generated comments into our own taxonomy. 
Figures~\ref{fig:rq2-part1} and \ref{fig:rq2-part2} present the distribution of comment types by LLM Reviewer and human reviewers for both Atlassian's internal and OSS projects, respectively.




\textbf{\underline{Results.}} 
\textbf{For Atlassian projects, the LLM reviewer generated a higher proportion of comments related to bugs and maintainability, but fewer comments focused on readability compared to human reviewers.}
When comparing between the LLM reviewer and human reviewers (see Figure~\ref{fig:rq2-part1}), we found that, the LLM reviewer generated comments related to code bugs more often than human reviewers did  (18.2\% [$\frac{128}{702}$] vs. 6.5\% [$\frac{30}{465}$]), and also generated more maintainability-related comments (26.8\% [$\frac{188}{702}$] for LLM vs. 19.1\% [$\frac{89}{465}$] for humans), representing approximately 2.8-fold and 1.4-fold increases, respectively. 
Conversely, although readability is the most frequently comment type for both groups, human reviewers emphasized it more than the LLM reviewer (46.0\% [$\frac{214}{465}$] vs. 33.2\% [$\frac{233}{702}$]).



\textbf{For OSS projects, the LLM reviewer generated more maintainability-related comments but fewer design-related comments than human reviewers did.}
As shown in Figure~\ref{fig:rq2-part2}, maintainability is the most common comment type from the LLM reviewer, comprising 37.9\% [$\frac{476}{1,256}$] of its comments, compared to 23.6\% [$\frac{236}{1,000}$] from human reviewers.
The LLM reviewer generated less no issue comments than human reviewers, with only 0.6\% falling into no-issue compared to 9.0\% for humans.
This could be because developers use code reviews for identifying issues as well as a communication channel to  steer the discussion.

These differences across Atlassian's internal and OSS projects highlight that the LLM reviewer and human reviewers exhibit distinct strengths and weaknesses depending on the project context.
This is likely because the nature of code review practices vary across different organizations, codebase familiarity, and project goals. 
For example, in Atlassian's internal projects, the LLM reviewer may be more effective at systematically identifying bugs and maintainability issues due to access to structured guidelines and consistent coding standards, whereas human reviewers might leverage their deeper contextual knowledge to focus on readability and nuanced design concerns. 
In OSS projects, the diversity of contributors and less standardized practices can lead the LLM reviewer to emphasize maintainability, while human reviewers—often more familiar with the project's architecture and community norms—may provide more design-related feedback. 

\subsection*{\textbf{(RQ2) \RQtwo}}
\begin{figure}[t]
    \centering
    \includegraphics[width=\linewidth]{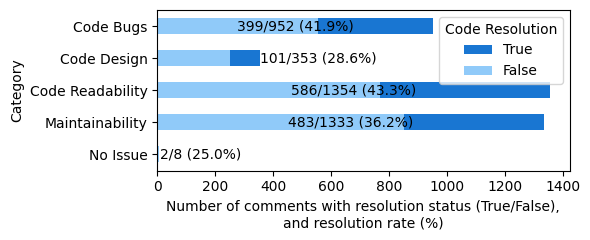}
    \caption{(RQ2) The percentage of code resolution of the LLM-generated comments for each comment type at Atlassian.}
    \label{fig:rq3-resolution}
\end{figure}

\textbf{\underline{Approach.}} To answer this RQ, we aim to examine the association between the comment types and the resolution rate.
To examine whether the generated comments are resolved or not, we conduct an online evaluation.
Similar to RQ1, we employed the RovoDev Agent in the Atlassian's internal projects. 
Since code review practices are matured and changed overtime, we opted to focus on the most recent code review comments, spanning across a two-month period (June-July) in 2025.
Given a large amount of LLM-generated comments at Atlassian, we randomly selected 4,000 LLM-generated comments that were associated with 3,746 pull requests, spanning across 1,007 repositories.
Similar to RQ1, for each comment, we used our LLM Judge (Section III.C) to automatically classify the comment type.
A successfully resolved code review comment is determined 
similar to prior work~\cite{BosuGB15,RahmanRK17}.
We consider \emph{a given code review comment to be resolved} if a subsequent commit modified the exact line where the comment was placed.
Then, for each comment type, we measure the code resolution rate as the percentage of the resolved comments to the total of the LLM-generated comments.
Finally, we present the percentage of code resolution of the LLM-generated comments for each comment type (see Figure~\ref{fig:rq3-resolution}).


 
\textbf{\underline{Results.}} \textbf{Readability, bug, and maintainability-related comments exhibited higher resolution rates compared to those focused on code design.}
We observe that comments discussing code readability achieved the highest resolution rate at 43.3\% (586/1,354), followed by code bugs at 41.9\% (399/952), and maintainability at 36.2\% (483/1,333).  
In contrast, code design had a lower resolution rate of 28.6\%.  

Nevertheless, many of the LLM-generated comments are not resolved by developers (60\%-70\%). 
Such unresolved comments could be due to the lack of clarity, relevancy, and simplicity of the comments. 
Comments that are less clear or not directly relevant to the code being reviewed are less likely to be acted upon. 
Additionally, we observed that code design issues often involve more complex, architectural considerations that require deeper understanding and broader changes, making them harder to resolve quickly or automatically. 
As a result, developers may hesitate to address these suggestions without further discussion or context, leading to lower resolution rates for design-related comments.


\section{Discussion}

\textbf{The LLM reviewer and human reviewers are complementary, as they focus on different types of code review comments.}   
Our RQ1 showed that the LLM reviewer and human reviewers exhibited distinct strengths and weaknesses depending on the project context. 
For instance, in the Atlassian's projects, the LLM reviewer generated a higher proportion of comments related to bugs and maintainability, but fewer comments focused on readability compared to human reviewers. 
Despite these differences in focus, our RQ2 showed that many of LLM comments were resolved in practices, indicating the value of the LLM comments. 

Our results show that LLM reviewers provide different yet acceptable feedback compared to human reviewers. 
This supports prior work arguing that evaluation should consider comment types rather than relying solely on lexical similarity metrics like BLEU, which may be insufficient \cite{tufano22, lin24, tufano21, tufano_strengths, hong2022commentfinder}. Future research should explore alternative evaluation aspects, e.g., comment type coverage and actionability, to guide the development of metrics beyond using human-written comments as the sole ground truth.


\textbf{The LLM reviewer should  balance the distribution of comment types and improve the clarity and relevancy.}
RQ1 showed that the LLM reviewer tends to generate fewer bug-related comments than other types (e.g., 17.8\% less than maintainability for OSS projects and 15\% less than code readability for Atlassian's internal projects).
However, many prior studies \cite{bosu2015characteristics, TurzoB24, BosuGB15} showed that code bugs-related comments are among the most useful for human reviewers. 
Our RQ2 findings support this, showing that bug-related comments have the second-highest resolution rate (5.7\% higher than maintainability), indicating that developers are more likely to address them.
Nonetheless, we observed that unresolved comments could be due to the lack of clarity, relevancy, and simplicity of the
comments.
Hence, balancing the distribution of comment types and improving the clarity and relevancy of review comments may improve the effectiveness of LLM-powered code review tools. 
Future work should address this limitation and explore strategies to increase the practical impact of automated code review.


\section{Threats to the Validity}


\textbf{Threats to Construct Validity} may arise from the use of GPT-4.1 as an annotator. The classification of comment types could vary if different large language models (LLMs) are employed. However, we have conducted experiments with other LLMs (e.g., Anthropic Claude), and observed that the results remain consistent.


\textbf{Threats to Internal Validity} may arise from the inclusion of noisy LLM-generated comments in the analysis. 
To mitigate this, we focused exclusively on comments anchored to specific code lines, as these are likely to be specific and directly related to the code. 
Since RovoDev Agent is proprietary and its underlying architecture and prompt are confidential, full details cannot be disclosed.

\textbf{Threats to External Validity} concern the generalizability of our findings.
Our results are limited to the studied contexts (13 open-source and 1,007 Atlassian's internal projects) and to comments generated by a specific LLM.
Additionally, we used one internal LLM reviewer, i.e., RovoDev Agent since the study involved proprietary data (i.e., internal projects).
Therefore, the findings may not generalize to other projects, LLMs, or development environments.
Future work could explore the use of other LLMs to other projects.



\section{Conclusions}

In this paper, we investigate the types of review comments written by humans and LLMs, and the types of LLM comments that are most frequently resolved by developers.
Through a study across open-source and Atlassian's internal projects, we found that LLM and human reviewers focus on different types of code review comments, highlighting their complementarity.
Comments related to readability, bugs, and maintainability had higher resolution rates than those focused on code design. However, bug-related comments remain underrepresented in LLM comments.
To enhance the practical effectiveness of LLM-powered code review tools, future work should 1)
explore alternative evaluation aspects beyond using human-written comments as ground truth; 2) improve the balance of comment types, with  an increase focus on bug-related comments; and 3) improve the  clarity and relevancy of the LLM-generated comments.

\section*{Acknowledgment \& Disclaimer}
We acknowledge the AutoReview (DevAI) team, especially Yaotian Zou, Mohanish Ranade, Andy Wong, Michael Gupta, and Cameron Gregor for their support.

\textbf{Disclaimer.} The perspectives and conclusions presented in this paper are solely the authors' and should not be interpreted as representing the official policies or endorsements of Atlassian or any of its subsidiaries and affiliates. Additionally, the outcomes of this paper are independent of, and should not be constructed as an assessment of, the quality of products offered by Atlassian.

\bibliographystyle{IEEEtran}
\bibliography{references}

\end{document}